\documentclass[referee,epsf,rotate,psfig]{aa}
\usepackage{psfig,rotate}
\title{Recent Timing Studies on RXTE Observations of 4U 1538-52}
\author{A. Baykal$^1$, S.\c{C}. \.{I}nam $^2$ and E. Beklen $^{1,3}$ \\ $^1$ Physics Department,\\ Middle East Technical University, 06531 Ankara, Turkey \\ altan$@$astroa.physics.metu.edu.tr \\ $^2$ Department of Electrical and Electronics Engineering,\\ Ba\c{s}kent University, 06530 Ankara, Turkey \\ inam$@$baskent.edu.tr \\ $^3$ Physics Department, \\ S\"{u}leyman Demirel University, 32260, Isparta, Turkey \\ elif$@$astroa.physics.metu.edu.tr}
\institute{}
\date{}
\begin{document}
\abstract{The high mass X-ray binary pulsar 4U 1538-52 was observed between
July 31 and August 7, 2003. Using these observations, we  
determined new orbital epochs for both circular and elliptical orbit 
models. The orbital epochs for both orbit solutions 
agreed with each other and yielded an orbital period derivative 
$\dot{P} / P = (0.4 \pm 1.8 ) \times 10^{-6}$ yr$^{-1}$.
This value is consistent with the earlier 
measurement of $\dot{P} / P = (2.9 \pm 2.1 ) \times 10^{-6}$ yr$^{-1}$ 
at the $1 \sigma$ level and gives only an upper limit to the 
orbital period decay.
Our determination 
of the pulse frequency  showed that the source spun up at an average rate 
of $2.76 \times 10^{-14}$ Hz sec$^{-1}$ between 1991 and 2003.

}
\maketitle
\section {Introduction}
The source 4U 1538-52 was discovered using the Uhuru satellite 
(Giacconi et al. 1974). X-ray pulsations with a period of 529 seconds 
were detected from two independent satellite experiments: 
Ariel 5 (Davidson 1977) and OSO-8 (Becker et al., 1977).
OSO-8 observations also revealed a clear orbital modulation at 
3.7 days and evidence of an eclipse lasting $\sim$0.51 days.
The optical companion of 4U 1538-52 was identified to be the B0 I giant 
QV Nor (Parkes, Murdin \& Mason 1978).
 BATSE observations of this source permitted the construction
of long-term pulse frequency and intensity histories (Rubin, Finger, Scott et
al. 1997). In the pulse frequency history, Rubin et al. (1997) found
short-term pulse frequency changes of either sign, and a 
power density spectrum 
of fluctuations of the pulse frequency derivative that is consistent 
with white torque noise on timescales from 16 to 1600 days.

From RXTE observations,
 Clark (2000) obtained new orbital
 parameters of the source which provided marginal evidence of 
orbital decay, i.e., they found
$\dot{P}_{orb}/P_{orb}=(-2.9\pm 2.1)\times 10^{-5}$ yr$^{-1}$. 
In this work, we present new orbital epoch and pulse frequency measurements 
based on our analysis of archival RXTE observations of 4U 1538-52.

\section {Observations}

 The observations of 4U 1538-52 took place between July 31 and 
 August 7, 2003 (MJD 52851 - 52858) and accumulated a total nominal 
 exposure of $\sim 75$ ksec. The results presented 
 here are based on data collected with the Proportional Counter Array (PCA, 
Jahoda et al., 1996). The PCA instrument consists of an array of five
 collimated xenon/methane multianode proportional counters. The total effective
 area is approximately 6250 cm$^{2}$ and the field of view is $\sim 1 ^{0} 
 $FWHM. The nominal energy range of the PCA extends from 2 to 60 keV. 

\section {Determination of Orbital Epoch and Pulse Frequency}

\begin{figure}[h]
\begin{center}
\psfig{file=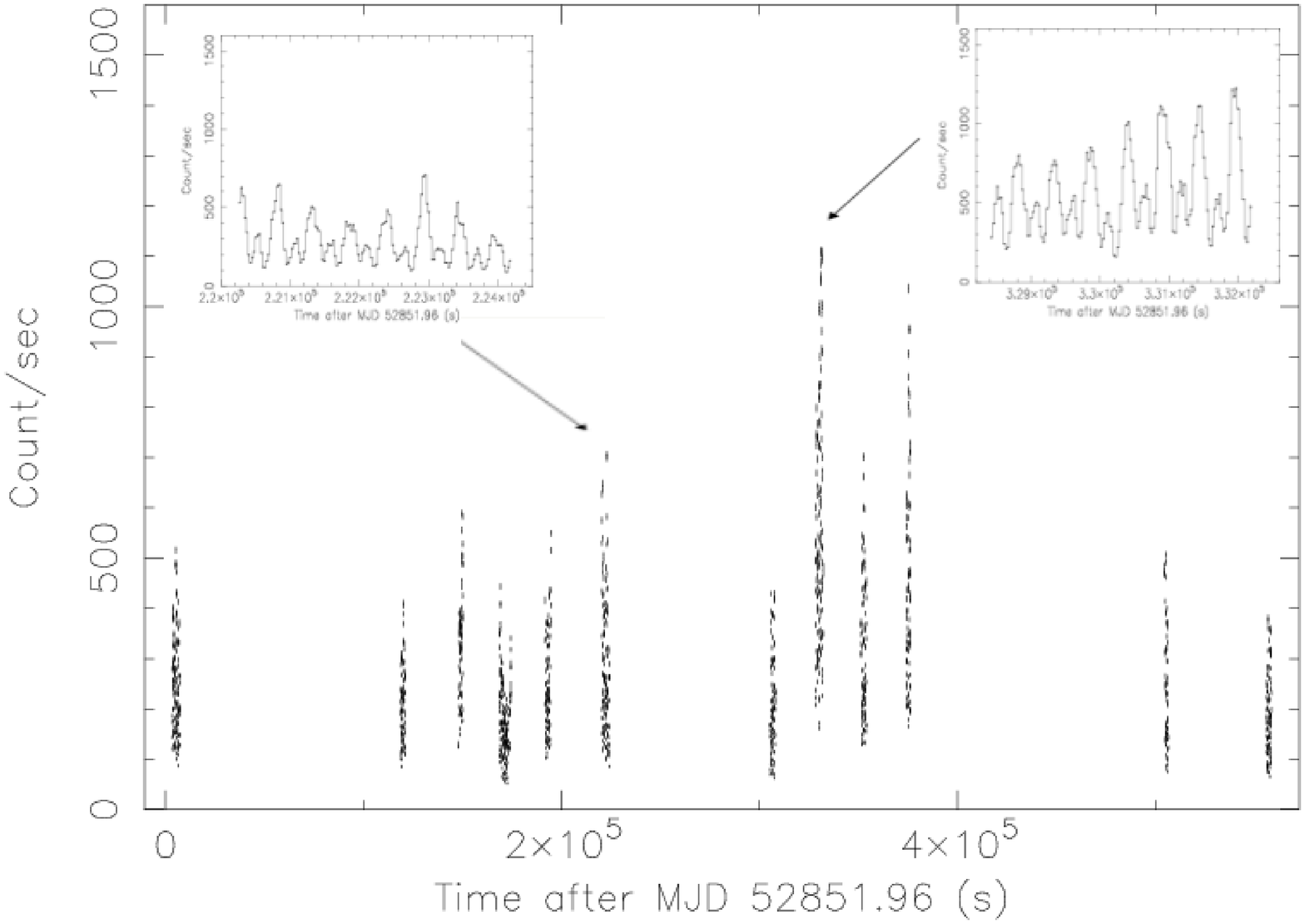,width=14.5cm,angle=-90}
\small{Fig. 1 --  2-30 keV RXTE-PCA light curve of 4U 1538-52 
between July 31 and August 7, 2003. Two 26s binned $\sim 3$ksec samples 
of this light curve corresponding to single RXTE orbits are presented on the 
upper left and the upper right.}
\end{center}
\end{figure}

 Background light curves and X$-$ray spectra were generated by using background 
estimator models based on the rate of very large events (VLE), 
detector activation, and cosmic X$-$ray {\bf{background}}. 
 The background light curves were subtracted from the source light curve
obtained from the binned Good Xenon data. {\bf{For 4U 1538-52, X-ray emission coming from the galactic ridge is only a few percent of the total X-ray emission which should not affect our timing analysis (see Makishima et al. 1987), so we did not include an estimation of the galactic ridge emission in our analysis.}} In Figure 1, we present 
the background subtracted light curve. Although the number of 
active PCUs varied from 1 to 4 during the observations, 
Figure 1 shows count rates 
adjusted as if 5 PCUs had been active using the "correctlc" tool in 
HEASOFT 6. For the timing analysis, we corrected 
the light curve to the barycenter of the solar system.  
We also corrected this barycentered 
light curve for binary orbital motion using  
both circular and elliptical orbital models given in Table 1 (see also 
Clark (2000)). Then a template pulse profile
was created by folding the light curve into one master profile. 
Pulse profiles were also made from each of the 12 independent 
RXTE observations seen in Figure 1.

In order to find the pulse frequency and a new orbital epoch, we 
obtained
12 pulse arrival times through the $\sim 2$ binary orbits
using {\bf{a cross-correlation technique}}.  
In the pulse timing analysis, we used the method of
harmonic representation of pulse profiles, which was proposed by
Deeter $\&$ Boynton (1985). In this method, the pulse profiles 
for each orbit and the master profile are expressed in terms of harmonic 
series. We used 10-term unweighted harmonic series to
cross-correlate
 the template pulse profile with the pulse profiles for each 
RXTE observation. The maximum value
of the cross-correlation is analytically well-defined and does not depend
on the phase binning of the pulses. 

The source 4U 1538-52 has a variable pulse profile which
affects the pulse timing. In order to estimate
 the errors in the arrival times,
the light curve of each RXTE observation was
 divided into approximately 4-5 equal 
subsets and new arrival times were estimated.
The standard deviation of the arrival {\bf{times}} obtained
from each subset of the observation was taken to be the 
uncertainty in the arrival time for that observation. 

{\bf{Arrival time delays}} may arise from the
change of the pulse frequency during the observation (or
intrinsic pulse frequency derivative) and from the {\bf{differences
between the assumed and actual orbital and pulse parameters}} 
(Deeter, Boynton and Pravdo 1981),
\begin{equation}
\delta \phi = \phi_{o} + \delta \nu (t-t_{o})
+  \frac{1}{2} \dot \nu (t-t_{o})^{2}
- \delta \nu \frac{2\pi \delta T_{\pi/2}}{P_{orbit}} \frac{asin i}{c} cos l_{n}
\end{equation}
where $\delta \phi $ is the pulse phase offset
deduced from the pulse
timing analysis,  $t_{o}$ is the mid-time of the observation, 
$\phi_{o}$ is the  residual phase offset at t$_{o}$,
$\nu$ is the pulse frequency at time $t_0$, 
$\dot{\nu}$ is the pulse frequency derivative of the source,
$T_{\pi/2}$ is the orbital epoch when the mean orbital
longitude is equal to 90 degrees, $P_{orbit}$ is the orbital period 
and $l_{n}=2\pi (t_{n}-T_{\pi/2})/P_{orbit} +\pi/2$ is the mean orbital
longitude at $t_{n}$.
 Corrected values of 
orbital {\bf{and pulse}} parameters 
$\delta \nu $, $\delta T_{\pi /2}$ and $\dot{\nu}$
 were estimated from 
the fits of above expression to pulse phase residuals.

{\bf{Table 1 
presents the result of fits for both orbital and pulse parameters.}} 
We also reestimated the orbital epochs by varying the 
projected orbital radius $(a_{x}/c) sin i$ 
in the range of its uncertainty ($\pm 1 \sigma$) and
found that the resulting orbital epochs are consistent with the best fit 
value at the 1$\sigma $ level. The error in the orbital epoch 
due to an error in $(a_x/c)\sin i$ may 
also be expressed (Deeter, Boynton and Pravdo 1981) as 
\begin{equation} 
\sigma _{T_{\pi/2}}=\frac{P_{orb}\sigma _{(a_{x}/c sin i})}
{2\pi a_{x}/c sin i} \sim  0.0073 days
\end{equation}
where $\sigma _{a_{x}/c sin i}$ is the uncertainty in the 
projected orbital radius. This value is small relative to our 
error estimate for the orbital epoch (see Table 1).
 In Figure 2, we present the arrival 
time delay, the best fit elliptical orbit model, 
and arrival time residuals.

As seen from Table 1, the 
orbital epochs for circular and elliptical orbital models agree with
each other at the 1 $\sigma $ level. In order to check our 
technique, we extracted observations of 4U 1538-52 done in
1997 (MJD 50449.93-50453.69) and estimated orbital epochs for 
those observations. The results agreed with the orbital epochs 
given by Clark (2000). 

In Table 2, we present the orbital epoch measurements from  
different observatories and orbital cycle number (n).
In Figure 3, we present observed minus calculated values 
of orbital epochs
($T_{\pi/2}-n<P_{orbit}>-<T_{\pi/2}-n<P_{orbit}>>$) relative to the 
constant orbital period ($<P_{orbit}>=3.7228366$ days). 
A quadratic fit to the epochs from all experiments yielded 
an estimate of the rate of period change  
$\dot P_{orb}/P_{orb} =(0.4 \pm 1.8) \times 10^{-6}$ yr$^{-1}$. 
In Figure 4, we display the long-term pulse frequency history of the source. 

\begin{table}[hb]
\caption{Orbital Parameters of 4U 1538-52}
\begin{center}
\begin{tabular}{c | c c} \hline
Parameter &  Elliptical Orbit & Circular Orbit \\ \hline
T$_{\pi/2}$ Orbital Epoch (MJD) & 52855.0421$\pm$0.025 & 52855.0441$\pm$0.025 \\
a$_{x}$ sin i (lt-s)$^{a}$ & 56.6$\pm$0.7 & 54.3$\pm$0.6 \\
e$^{a}$                          & 0.174$\pm$0.015   &              \\
$\omega$$^{a}$(deg)              & 64$\pm$ 9         &               \\
$P_{orbit}$$^{a}$              &3.7228366$\pm$0.000032 &3.7228366$\pm$0.000032      \\
Epoch (MJD)     & 52855.0585$\pm$0.025 &52855.0585$\pm$0.025  \\
P$_{pulse}$ (s) & 526.8551$\pm$0.016 &  526.8535$\pm$0.013 \\
$\dot \nu$ (Hz s$^{-1}$)  &  (2.838$\pm$4.124)$\times$ 10$^{-13}$ & 
(2.241$\pm$ 2.764)$\times$ 10$^{-13}$ \\
reduced $ \chi^{2}$  & 1.44 & 1.0 \\ \hline
\end{tabular}
\end{center}
$^{a}$ Taken from Clark (2000)
\end{table}

\begin{table}[hb]
\caption{Orbital epochs by pulse timing analysis}
\begin{center}
\begin{tabular}{c c c c} \hline
Experiment & Orbit Number & Orbital Epoch (MJD) & Reference \\ \hline
OSO 8 & -1128 &      43015.800$\pm$ 0.1  &  Becker et al., 1977 \\
Tenma & -457  &     45517.660$\pm$ 0.050 &  Makishima et al., 1987 \\
Ginga &  0    &    47221.474$\pm$ 0.020  &  Corbet et al., 1993\\
BATSE &  370  &    48600.979$\pm$ 0.027  &  Rubin et al., 1997\\
BATSE &  478  &    49003.629$\pm$ 0.022  &  Rubin et al., 1997\\
BATSE &  584  &    49398.855$\pm$ 0.029  &  Rubin et al., 1997\\
BATSE &  691  &    49797.781$\pm$ 0.022  &  Rubin et al., 1997\\
RXTE  &  866  &    50450.206$\pm$ 0.014  &  Clark 2000 \\
RXTE  & 1511  &     52855.0421$\pm$ 0.025& present work\\ \hline
\end{tabular}
\end{center}
\end{table}

\begin{figure}[h]
\begin{center}
\psfig{file=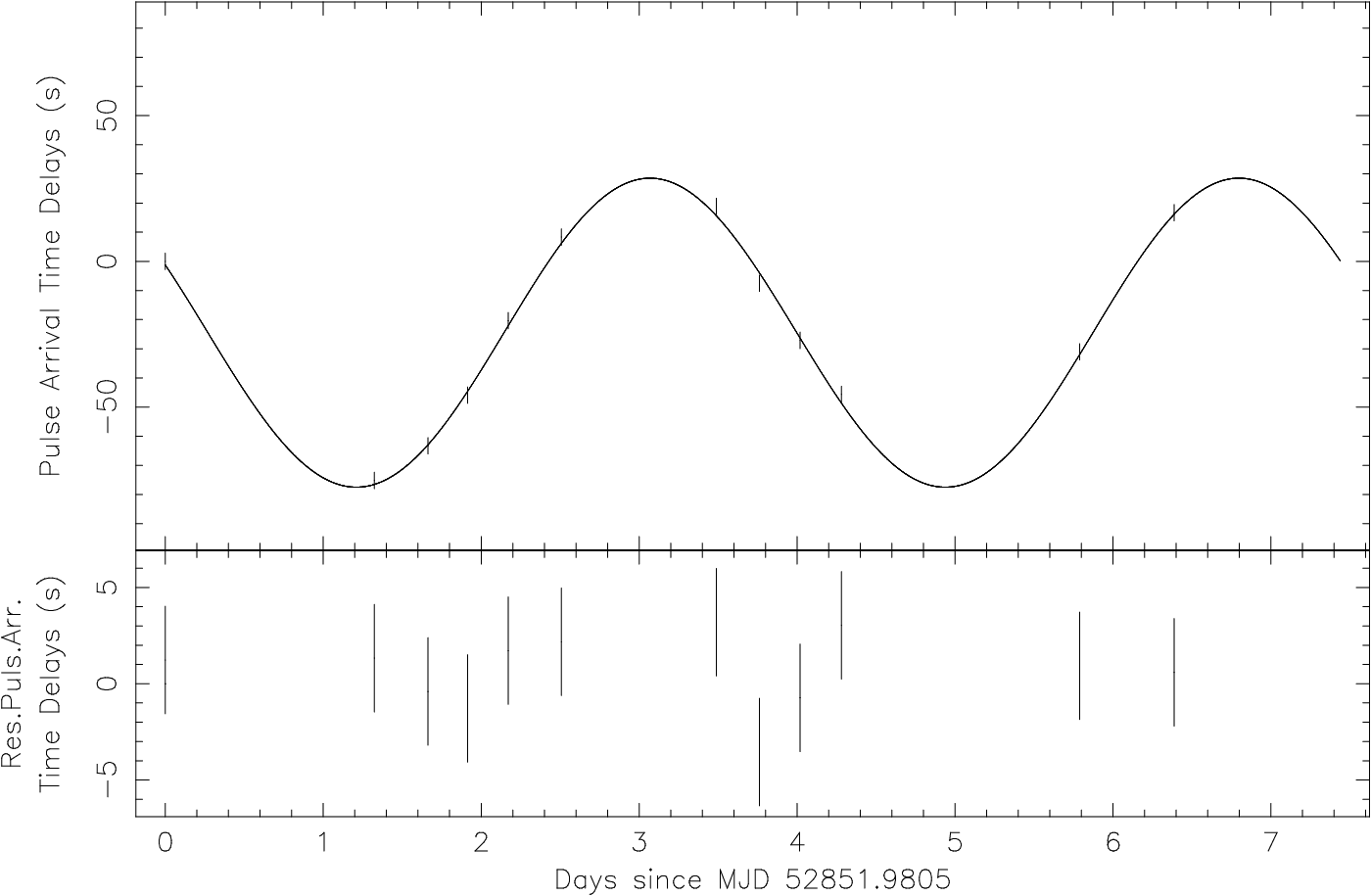,height=12cm,width=14cm,angle=-90} 
\small{Fig. 2 -- {\bf{(Top)}} Pulse arrival time delays 
 and best-fit
elliptical orbital model given in Table 1.
(Note that pulse profiles are obtained 
with respect to the reference time 52855.0585 MJD). 
{\bf{(Below)}} Residuals after removing best orbital model}.
\end{center}
\end{figure}

\begin{figure}[h]
\begin{center}
\psfig{file=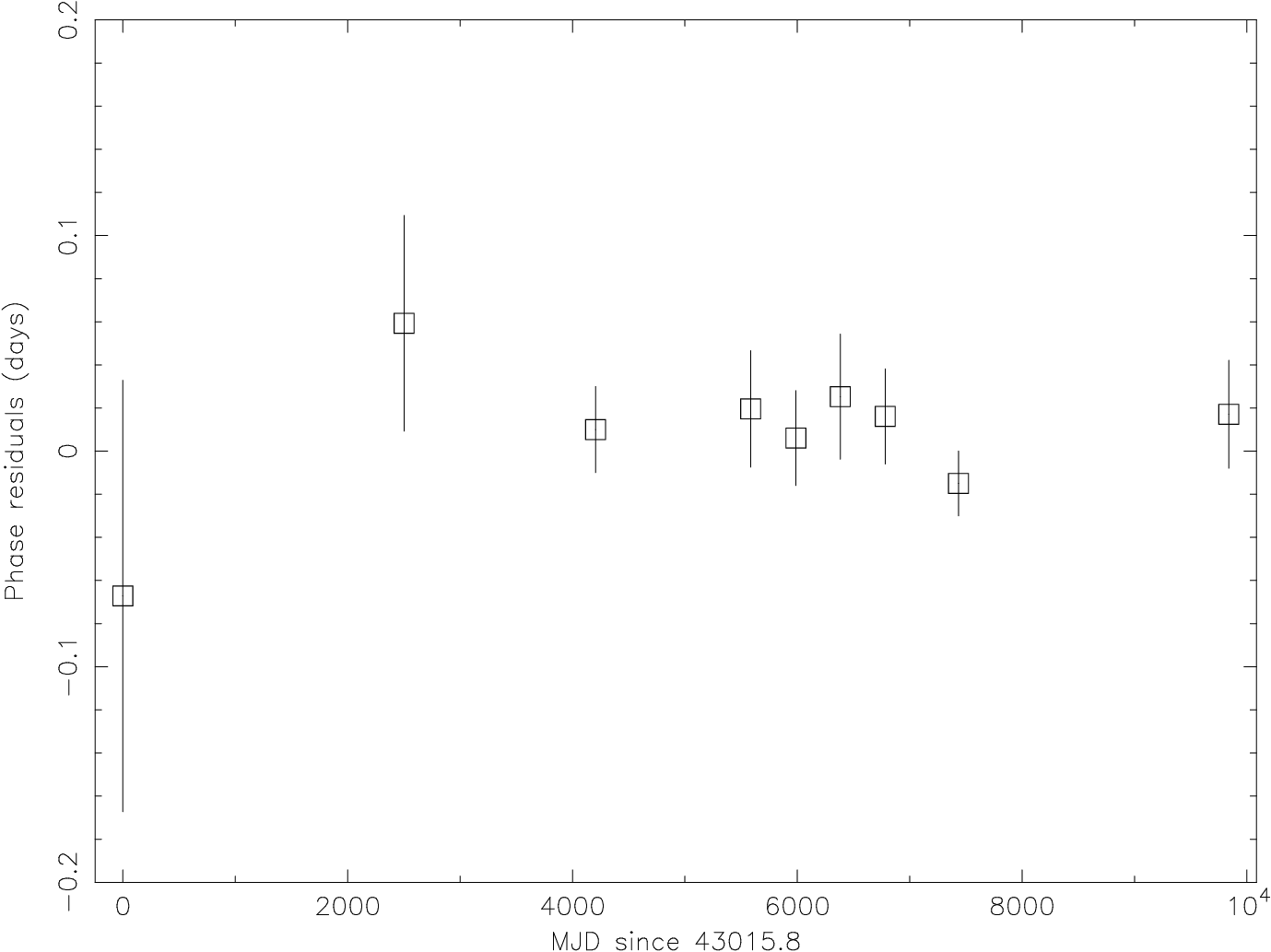,height=10cm,width=12cm,angle=-90} 
\small{Fig. 3 -- The phase residuals of orbital epoch for 4U 1538-52. 
The  orbital phases are estimated  relative the constant 
orbital period 
$(T_{\pi/2}-n<P_{orbit}>-<T_{\pi/2}-n<P_{orbit}>>)$, where n is the 
orbital cycle number (see Table 2). 
  The rightmost point
corresponds to the most recent RXTE observation of ID 80016.}
\end{center}
\end{figure}

\section{Discussion}

\begin{figure}[h]
\begin{center}
\psfig{file=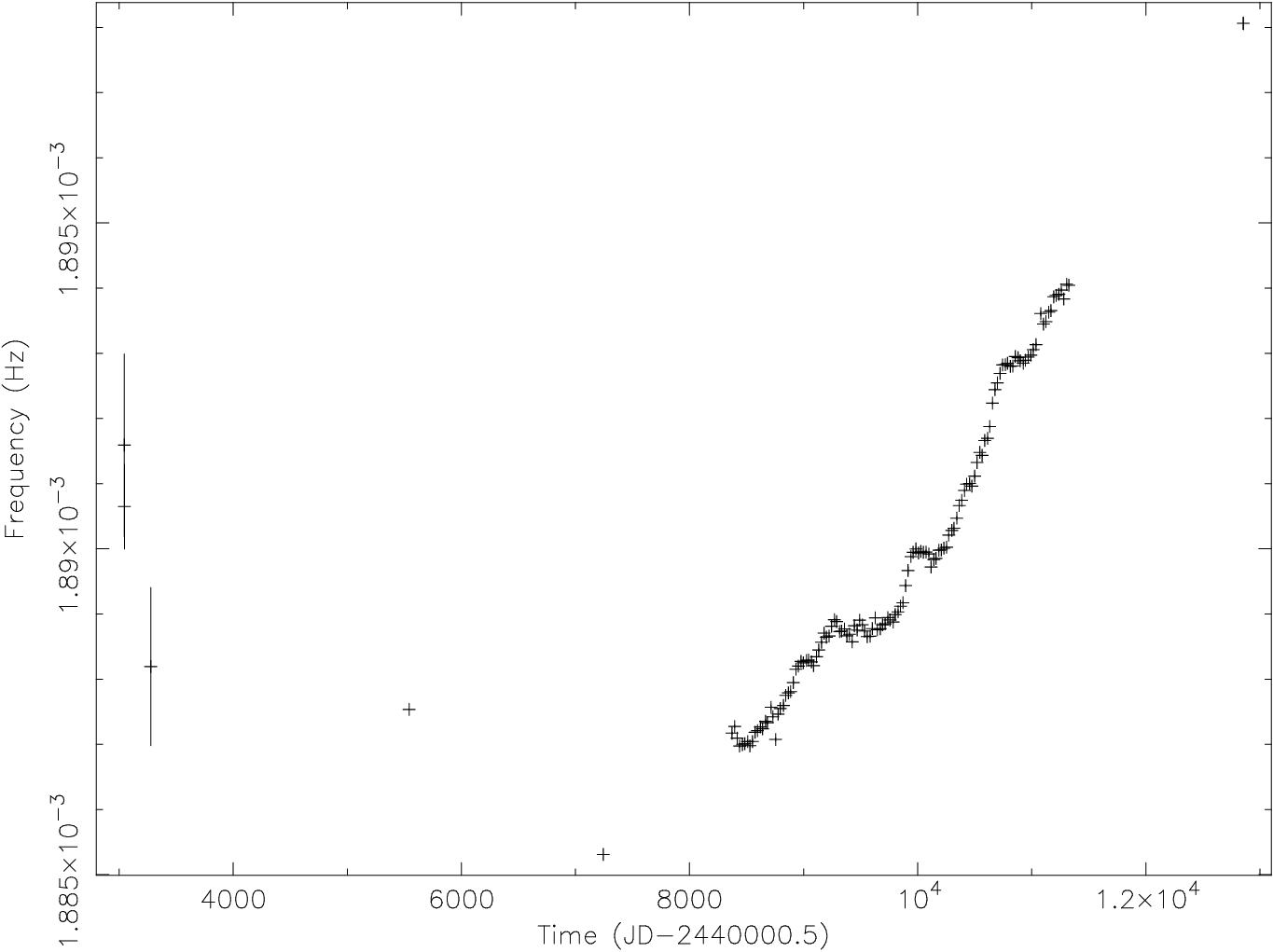,height=10cm,width=12cm,angle=-90} 
\small{Fig. 4 -- Pulse frequency history of 4U 1538-52. 
The rightmost point
corresponds to most recent RXTE observation of ID 80016.}
\end{center}
\end{figure}

Before CGRO observations, 4U 1538-52 had been found to have a long-term 
spin down trend. A linear fit to pre-CGRO pulse frequency history gives 
$\dot{\nu} /\nu \sim -8 \times 10^{-12} s^{-1}$ and a linear fit to 
CGRO and our RXTE result yields
 $\dot {\nu} /\nu \sim 1.45 \times 10^{-11} s^{-1}$. 
Rubin et al (1997) constructed the power spectrum of pulse frequency 
derivative fluctuations. Their analysis showed that the pulse frequency 
derivative fluctuations can be explained on timescales from 16 to 1600 days 
with an average white noise strength of (7.6 $\pm$ 1.6)$\times 10^{-21}$
(Hz s$^{-1}$)$^{2}$ Hz$^{-1}$.  A random walk in pulse frequency 
(or white noise in pulse frequency derivative) can be explained 
as a sequence of {\bf{steps in pulse frequency with an RMS value 
of $<(\delta \nu ^{2})>$}} which occur 
at a constant rate R. Then the RMS variation 
of the pulse frequency scales with elapsed time $\tau$
as $<(\Delta \nu) ^{2}> = R <(\delta \nu) ^{2}> \tau$ (Hz),  
where $S= R <\delta \nu ^{2}>$ is defined as noise strength.
Then, RMS scaling for the pulse frequency derivatives 
can be obtained as 
$<(\Delta \dot \nu) ^{2}>^{1/2} = (S/\tau )^{1/2} {\rm{Hz.s}}^{-1}$. 
As seen from Table 1, in our fits, upper limits on 
intrinsic pulse frequency derivatives are 7-10 times higher than 
the long-term spin up rates. If white noise in the pulse 
frequency derivative can be interpolated to a few days, {\bf{then the upper limit on the change of}} frequency derivative obtained from a $\sim 1$ week observation should 
typically have a magnitude that can be estimated from  
 $<(\Delta \dot \nu) ^{2}_{week}>=
  <(\Delta \dot \nu) ^{2}_{1600 days} \times 15^2 $ {\bf{This value is 15/7 - 15/10 times higher than the measured upper limit values. Therefore the measured upper limits on the intrinsic pulse frequency derivatives for 1 week are consistent with the values from the extrapolation of the power spectrum within a factor of a few.}} 

Previous marginal measurement of change in the orbital period, 
was (-2.9 $\pm$ 2.1)$\times 10^{-6}$ yr$^{-1}$ (Clark 2000), and our new 
value for the orbital period change,
$\dot{P}/P = (0.4 \pm 1.8 ) \times 10^{-6}$yr$^{-1}$,
are consistent with zero. 
These two measurements are consistent with each other in 
$1 \sigma $ level.

In most of the X-ray binaries with accretion powered pulsars,
 the evolution of the orbital period seems to be too slow to be detectable.
 Yet there are still some such systems in which this evolution was 
measured and $\dot{P}/P$ were reported.
 These systems include Cen X-3 with
(-1.8 $\pm$ 0.1)$\times$10$^{-6}$ yr$^{-1}$
(Kelley et al. 1983; Nagase et al. 1992),
 Her X-1 with (-1.32 $\pm$ 0.16)$\times$10$^{-8}$ yr$^{-1}$ (Deeter et al. 1991), 
 SMC X-1 with (-3.36 $\pm$ 0.02) $\times$10$^{-6}$ yr$^{-1}$ (Levine et al. 1993),
Cyg X-3 with (1.17 $\pm$ 0.44) $\times$10$^{-6}$ yr$^{-1}$ (Kitamoto et al. 1995), 
4U 1700-37 with (3.3 $\pm$ 0.6) $\times$10$^{-6}$ yr$^{-1}$ (Rubin et al. 1996),
and LMC X-4 with (-9.8 $\pm$ 0.7)$\times$10$^{-7}$ yr$^{-1}$ 
(Levine et al. 2000). Change in the orbital period of Cyg X-3 was associated 
with the mass loss rate from the Wolf-Rayet companion star. For 4U 1700-37,
 the major cause of orbital period change was also thought to be
 mass loss from the companion star. For Her X-1, mass loss and mass
 transfer from the companion were proposed to be the reasons of the change in 
 the orbital period of the system. 

On the other hand, for the high mass X-ray binary systems
 Cen X-3, LMC X-4 and SMC X-1, the major cause of change
 in the orbital period is likely to be tidal interactions
 (Kelley et al. 1983; Levine et al. 2000; Levine et al. 1993).
 For these three systems, orbital period decreases
 (i.e. derivative of the orbital period is negative).
 Our new measurement of orbital period change 
 ($\dot{P}/P$) gives the value of about $-10^{-6}$ yr$^{-1}$ which is
similar to the observed values of SMC X-1 and Cen X-3. Further observations 
can give further information about the orbital period change of this source.

{\bf{Acknowledgments}}

We thank anonymous referee for useful suggestions and comments.

\section*{References}

\noindent{Becker R.H., Swank J.H., Bold E.A., Holt S.S., Pravdo S.H., 
          Saba J.R., Serlemitsos P.J., 1977, ApJ. Lett., 216, L11} 

\noindent{Clark, G.W. 2000, ApJL, 542, 133}

\noindent{Corbet, R.H.D., Woo, J.W., Nagase, F. 1993, A\& A, 276, 52}

\noindent{Deeter, J.E., Boynton, P.E., Pravdo, S.H. 1981, ApJ, 247, 1003}

\noindent{Deeter, J.E., Boynton, P.E. 1985, in Proc. Inuyama Workshop on Timing Studies of X-ray Sources, ed. S. Hayakawa\& F.Nagase (Nagoya: Nagoya Univ.), 29}

\noindent{Deeter, J.E., Boynton, P.E., Miyamoto, S. et al. 1991, ApJ, 383, 324}

\noindent{Giacconi R., Murray S., Gursky H., Kellog E., Schreier E., Matilsky T., Koch D., Tananbaum H., 1974, ApJS, 27,37} 

\noindent{Jahoda, K., Swank, J.H., Giles, A.B., Stark, M.J., Strohmayer, T., Zhang, W., Morgan, E.H., 1996, Proc. SPIE, 2808, 59} 

\noindent{Kelley, R.L, Rappaport, G., Clark, G.W., Petro, L.D. 1983, ApJ, 268, 790}

\noindent{Kitamoto, S., Hirano, A., Kawashima, K. et al. 1995, PASJ, 47, 233}

\noindent{Levine, A., Rappaport, S., Deeter, J.E. et al. 1993, ApJ, 410, 328}

\noindent{Levine, A.M., Rappaport, S.A.,
 Zojcheski, G., 2000, ApJ, 541, 194}

\noindent{Makishima K., Koyama K., Hayakawa S., Nagase F., 1987, 
          ApJ., 314, 619}

\noindent{Nagase, F., Corbet, R.H.D., Day, C.S.R., et al. 1992, ApJ, 396, 147}

\noindent{Parkes, G.E., Murdin, P.G., Mason, K.O. 1978, MNRAS, 184, 73}

\noindent{Rubin, B.C., Finger, M.H., Harmon, B.A. et al. 1996, ApJ, 459, 259}

\noindent{Rubin, B.C.,
 Finger, M.H., Scott, D.M. et al. 1997, ApJ, 488, 413}

\end{document}